\documentclass[aps,prl,twocolumn,groupedaddress,showpacs]{revtex4}
\usepackage{graphicx}
\usepackage{latexsym}
\begin{document}

\title{Possible Field-Tuned SIT in High-T$_c$ Superconductors: Implications for Pairing at High Magnetic Fields}

\author{M. A. Steiner,$^1$  G. Boebinger,$^2$  A. Kapitulnik$^{1,3}$}
\affiliation{$^1$ Department of Applied Physics, Stanford University,
Stanford, CA 94305\\
$^2$ National High Magnetic Field Laboratory, Florida State University, Tallahassee, FL 32310\\
$^3$ Department of Physics, Stanford University, Stanford, CA 94305}

\date{\today}

\begin{abstract} 
The behavior of some high temperature superconductors (HTSC) such as $\rm La_{2-x}Sr_{x}CuO_{4}$ and $\rm Bi_{2}Sr_{2-x}La_xCuO_{6+\delta}$, at very high magnetic field, is similar to that of thin films of amorphous InOx near the magnetic field-tuned superconductor-insulator transition.  Analyzing the InOx data at high fields in terms of persisting local pairing amplitude, we argue by analogy that local pairing amplitude also persists well into the dissipative state of the HTSCs, the regime commonly denoted as the ``normal state" in very high magnetic field experiments.
\end{abstract}


\maketitle

In this paper we show that the behavior of some high temperature superconductors (HTSC) such as $\rm La_{2-x}Sr_{x}CuO_{4}$ \cite{ando1} and $\rm Bi_{2}Sr_{2-x}La_xCuO_{6+\delta}$ \cite{ando2}, at very high magnetic field is similar to that of thin films of amorphous indium oxide (InOx) near the magnetic field tuned superconductor-insulator transition (SIT). Comparing the details of the behavior of these two systems we conclude that the important ingredients in understanding the response of HTSC at  low temperatures to a high magnetic field are their quasi-two-dimensional nature and low superfluid density \cite{disorder},  which make the HTSC system highly susceptible to phase fluctuations \cite{emery}. Upon the application of magnetic field, we conjecture that the single Cu-O layer undergoes a magnetic-field-tuned superconductor to insulator transition at a critical field of order of the  mean-field upper critical field (denoted as $H_{c2}(0)$ \cite{hc2}). Further increasing the magnetic field beyond $H_{c2}(0)$  weakens the tendency towards a Bose-dominated insulating phase, very much what is observed in our InOx films \cite{uchida}.  We thus conclude that the regime denoted as the ``normal state" in the very-high field experiments of Ando {\it et al.} \cite{ando1,ando2} includes a large pairing susceptibility (i.e. local pair amplitude) that persists to fields as high as $\sim 3H_{c2}(0)$.

 Understanding the nature of the normal state of HTSC  is an important part of understanding the cause of superconductivity in these materials.  In particular, understanding the underlying normal state in the temperature range where the materials are superconducting has been attempted in many experiments.  For bulk low-temperature BCS superconductors (LTSC) this is easily accomplished by the application of a magnetic field larger than the upper critical field $H_{c2}$ \cite{hc2}, thereby breaking all pairs and usually revealing an underlying Fermi liquid state.  However, attempting such a procedure for HTSC has  proven difficult due to the intense magnetic fields required and the fragility of the vortex state.  While dissipation due to the melting or depinning of the vortices is easily achieved, recent experiments indicate that pairing and some form of local superconducting coherence  persist to very high magnetic fields \cite{ong1,ong2,capan}. 

\begin{figure}[ht]
\centering
\includegraphics[width=0.93\columnwidth]{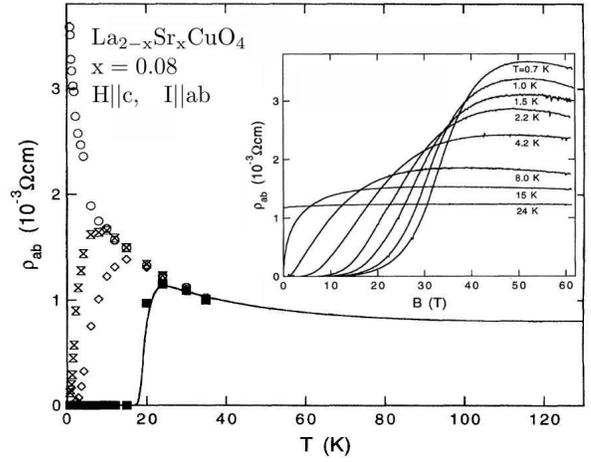}
\caption{ \footnotesize \setlength{\baselineskip}{0.8\baselineskip} A combination of Fig.~1(inset) and Fig.~2 (main panel) reprinted from Ando {\it et al.} \cite{ando1}.   $\rho_{ab}$ is the in-plane resistivity of  $\rm La_{1.92}Sr_{0.08}CuO_{4}$  in perpendicular magnetic fields of 0 (solid line), 10, 20, and 60 T.  The inset shows isotherms of magnetoresistance. Note the negative magnetoresistance above $\sim$50 T below $\sim$2 K.}
\label{ando}
\end{figure}

An influential early  attempt to reveal the underlying normal state of a prototype HTSC system was the seminal study of $\rm La_{2-x}Sr_{x}CuO_{4}$ (LSCO) by Ando {\it et al.} \cite{ando1}. Using  pulsed fields up to 60 T, these authors concluded that the resistivity  revealed at the high magnetic fields probes the behavior of  the normal state \cite{boeb}.  A logarithmic divergence with a large coefficient was measured at low temperatures throughout the underdoped regime, which prompted many explanations for this unusual ``normal state" \cite{boeb}.   However, not much attention was given to the peculiar result that a negative magnetoresistance  starts to develop at the highest fields and lowest temperatures, an effect clearly seen  in Fig.~\ref{ando} (which is adapted from the data of Ref.~\cite{ando1}). An intriguing observation is that the magnetoresistance peak at low temperatures occurs at a field of order  of the mean-field-zero temperature value of $H_{c2}$ that we would expect for this short coherence length ($\sim 20 \AA$) superconductor.  Taking this interpretation at face value, this means that at $0.7 K$, for $H \leq 52$ T,  the large resistance of the sample is dominated by weakly localized pairs, while above the peak for $H \geq 52$ T,  pairs start to dissociate at a faster rate, giving rise to a negative magnetoresistance as the system slowly approaches a state that does not support pairing.   This in turn suggests that the true superconducting transition is dominated by phase fluctuations. Unfortunately, much higher magnetic fields are needed to test whether the magnetoresistance continues to decrease to values of order of the resistance observed in zero magnetic field at higher temperatures.  However, we note that the proposed interpretation is independent of the pairing mechanism or symmetry (e.g. d-wave vs. s-wave) of the order parameter and thus can be tested on other quasi-two-dimensional superconductors with low superfluid density \cite{emery}. 

\begin{figure}[ht]
\centering
\includegraphics[width=0.93\columnwidth]{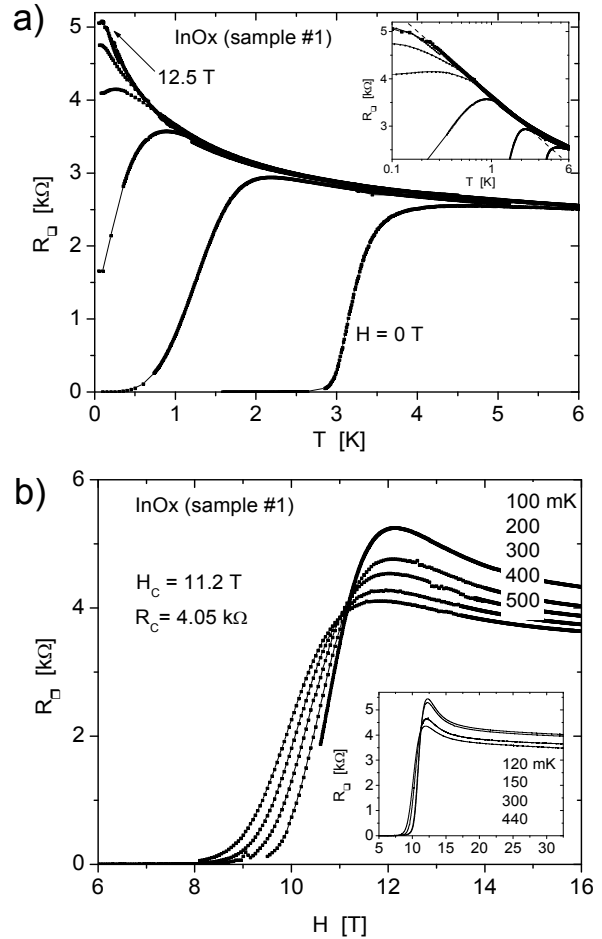}
\caption{ \footnotesize \setlength{\baselineskip}{0.8\baselineskip} (a) Resistive transitions of sample $\#$1 showing the resistance envelope. The fields are 0, 8.0, 10.5, 11.2, 11.5, 12.0 and 12.5 T. The inset shows the logarithmic behavior of this sample's envelope before a low temperature saturation (see text). The dashed line is a guide to the eye. (b) Magnetoresistance isotherms at five temperatures. The inset shows the magnetoresistance up to 32.5 T, where the peak is the highest at the lowest temperature of 120 mK. } 
\label{env}
\end{figure}

Indeed, such a magnetoresistance peak was first observed by Paalanen {\it et al.}  \cite{hebard2} in their investigation of the nature of the insulating state in InOx films near the SIT.  More recent  studies of SIT in thin superconducting films of amorphous MoGe \cite{nadya} and InOx \cite{sambandamurthy,myles} revealed a similar peak.  In particular Steiner and Kapitulnik \cite{myles} were able to tune the insulating behavior of their InOx films in a fashion that allows a  direct comparison with HTSC. Fig.~\ref{env} shows such results on an InOx sample that mimics the behavior of the LSCO shown in Fig.~\ref{ando}. We will return to the comparison between this InOx film and HTSC after we discuss the general properties of our InOx films and their SIT behavior.

%\section{Experimental}

Indium oxide is an amorphous low-carrier-density superconductor ($n\sim10^{20} - 10^{21}$ carriers/cm$^3$) \cite{kowal,hebard} and was used previously in the study of the SIT \cite{hebard1}.  Films of thickness 200 to 300 $\AA$ were prepared by electron beam evaporation of sintered In$_2$O$_3$ onto an acid-cleaned silicon nitride substrate \cite{kowal}.   $R_{\Box}$, $T_c$, and the overall SIT of the films were varied by adding oxygen during growth and by the subsequent careful annealing of the samples \cite{myles}.  The linear sheet resistance was measured using standard four-point lockin techniques at frequencies from 3 - 17 Hz and excitation currents 0.1 - 5 nA; the highest resistance sample was measured at dc. 

%\section{MAGNETORESISTANCE AND THE SIT}

To illustrate the controllability of the properties of the InOx films as pertain to their insulating behavior we show in this paper  three films (of the dozen or so films prepared and measured) of different normal-state $R_\Box$ and thickness for which the resistance was measured down to 50 mK and at magnetic fields up to 16 T.  The sample which is of particular interest to us for a direct comparison with HTSC  was also measured at fields up to 32.5 T.  Except for the very low fields,  for all samples, we observe a behavior similar to that shown in Fig.~\ref{env}a, where the data is dominated by the presence of a resistance envelope. The curves at progressively higher magnetic fields follow the envelope to lower temperatures before they deviate toward zero resistance.   From the lower field portion we determine the mean-field upper critical field, a procedure described in detail elsewhere \cite{myles}.  The  WHH \cite{whh} determination of $H_{c2}(0)$  gives the values 13, 9.5 and 9 T for samples $\#$1, 2 and 3 respectively with a a zero temperature coherence length  $\xi(0) \approx 50 \AA$ \cite{wl}.
   
\begin{figure}[ht]
\centering
\includegraphics[width=0.93\columnwidth]{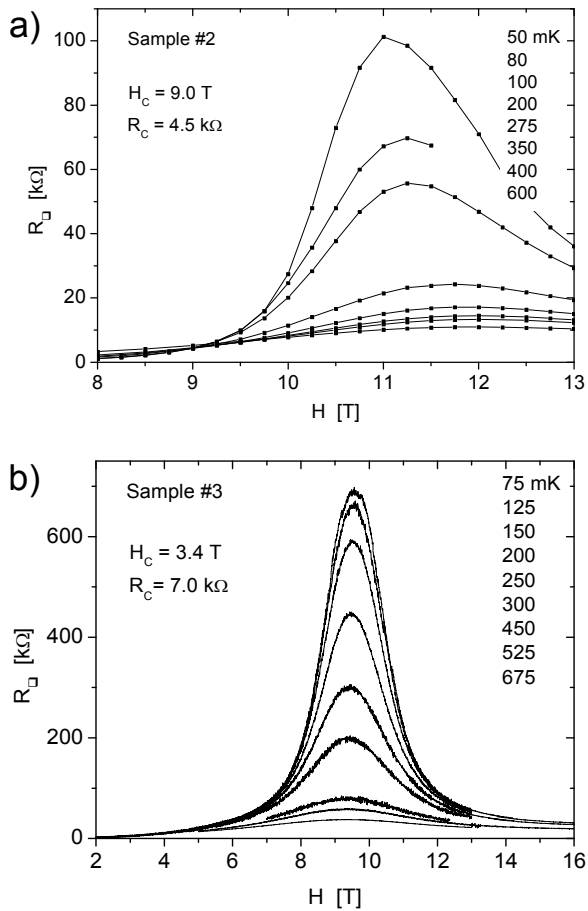}
\caption{ \footnotesize \setlength{\baselineskip}{0.8\baselineskip} Magnetoresistance isotherms for two samples with increased disorder relative to Fig \ref{env}. Crossing points at $\{ H_c,R_c\}$ mark the SIT. } 
\label{ins}
\end{figure}

The resistive envelope  presented here is qualitatively different from other reports on sputtered InOx \cite{hebard1} in which resistive curves appear to splay off from a common temperature, typically with wider transitions (see Fig.~1 in \cite{hebard1}). That pattern is more consistent with a granular system   where the overall behavior is dominated by the Josephson phase-coupling between grains of fixed $T_c$. By contrast, in our films most of the transition seems to be dominated by amplitude fluctuations, with a final phase-dominated SIT at low temperature.

At high fields and low temperatures the InOx films show signatures of a magnetic-field-tuned SIT \cite{hebard,yazdani,markovic,mk1}.  In Fig.~\ref{ins} we show the resistance at high fields for the other two films. Following a sharp rise in resistance the isotherms \emph{of all samples} go through a crossing point ($H_c,R_c$), a  signature of a zero temperature quantum phase transition. Close inspection of the crossing indicates that the transition broadens at very low temperatures, $\Delta H_c /\langle H_c \rangle \sim 1\%$, a feature previously discussed by Mason and Kapitulnik  \cite{mk1}.  Nevertheless, we can use $\langle H_c \rangle$ to scale the data using the usual one-parameter scaling form \cite{mpaf1}: $R(H,T) = R_c \mathcal{F}\left(\frac{H - \langle H_c\rangle}{T^{1/z\nu}}\right)$, similar to  \cite{hebard1}.  The fit gives $z\nu \simeq 1.3$, in agreement with other measurements on MoGe \cite{yazdani,mk1} and InOx \cite{hebard1}.  Above the crossing point  the resistance increases with decreasing temperature, marking an insulating phase at fields that are clearly lower than $H_{c2}(0)$.  $\langle H_c \rangle$ decreases  with disorder while $R_c$ increases, and is thus non-universal \cite{yazdani,mk1}.

The tendency towards a Bose-insulating phase does not persist and upon further increase of the magnetic field the isotherms reach a maximum and then start to decrease. The resistance at the maximum increases roughly exponentially with  temperature, with apparent saturation at the very lowest temperatures. The maximum for all samples is very close to the mean-field $H_{c2}(0)$. This is expected if we attribute the peak to a decrease in pairing susceptibility and a crossover of the system from being Bose-particle  to  Fermi-particle dominated. However, it does not appear, based on the available data range, that the resistance saturates at the highest accessible field.  Focusing on sample $\#$1 we measured it at fields up to 32.5 T (see inset of Fig.~\ref{env}b).   We note that the resistance at the lowest temperature (120 mK), at 32.5 T, is  $\sim$ 4 k$\Omega$, which is a factor of $\sim$ 1.7 higher than the zero field normal state resistance as extrapolated from the temperature dependent resistance above the transition.  Theoretically, for a simple metal with the characteristics of this film, we expect a small \cite{ziman,kapitza} $\Delta \rho/\rho_n \sim 10^{-3}$ \cite{kohler} and diminished weak-localization effects in such high fields \cite{ovadyahu}. \emph{Thus it appears that even at 32 T the film has not returned to its expected normal state}.

The insulating state above the crossing point is even more dramatic in the films with higher $R_\Box$, as shown in Figure \ref{ins}.  On the high field side of the peak, the isotherms all tend to decay to resistances $\leq$ 20 k$\Omega/\Box$, but still much above the normal state resistance. The increase in resistance in these stronger insulators is several orders of magnitude before the resistance starts to saturate at very low temperatures.  Taking fixed-field cuts through the isotherms \cite{sambandamurthy}, we find that the resistance increases as $R_\Box \sim e^{T_{A}/T}$ with a characteristic activation energy $k_BT_A$.  $T_A$ is of the order $T_{c0}$ (about 1 K) for the strongest insulator (sample $\#$3), while for the weak insulator (sample $\#$1) it is substantially lower; based on the limited temperature range of the data, we estimate $T_A$ to be at most $\sim$ 100 mK \cite{myles}.

%\section{Comparison with high-$\rm Tc$ superconductors}

Let us now  turn our attention to a comparison with the quasi-two-dimensional HTSC system. An envelope pattern similar to that seen in Fig.~\ref{env}a  was also observed by Ando \emph{et al.}  \cite{ando1} for the in-plane resistivity of underdoped $\rm La_{2-x}Sr_xCuO_4$ (LSCO) in a pulsed magnetic field. Their data,  (Fig.~\ref{ando}), shows an overall resistance envelope with curves at different fields splaying off at progressively lower temperatures. The inset to Fig.~\ref{ando} shows isotherms of the resistance which, at the lower temperatures, increase dramatically with field before peaking and slowly starting to decrease, similar to the behavior of the InOx shown in Fig.~\ref{env}b. Both the temperature and magnetic field scales of their experiment are much higher  than for the InOx films -- the envelope persists up to 50 T and the zero-field transition temperature was almost 20 K, but the qualitative resemblance to our data is clear. In fact, even quantitatively these two systems are similar, showing a logarithmic temperature dependence of the resistivity envelope $\Delta R /R= AR_\Box log(1/T)$ with a prefactor $A \gg pe^2/2\hbar \pi^2$ (for any reasonable $p$), indicating that the observed $log(1/T)$ behavior is not due to conventional weak localization.   Specifically for sample $\#$1 (see the inset of Fig.~\ref{env}a), $A \approx 1.3\times 10^{-4}$ $\Omega^{-1}$ as compared to an expected weak localization slope of $A \approx 2.4\times 10^{-5}$ (for $p=2$ \cite{ovadyahu}). The fact that the InOx films' resistance saturates at low temperatures while that of the HTSC does not may be the effect of the weak interlayer coupling. Indeed, a ground plane in proximity to a SIT in MoGe films was shown to overcome the saturation \cite{mk3}.   Additionally,  the InOx films show a sharply defined crossing point  in Fig.~\ref{env}b, while the $\rm La_{2-x}Sr_xCuO_4$ samples in Fig.~\ref{ando} do not exhibit a crossing point.  This might be attributed to the relatively high temperatures at which the HTSC data were taken, and the  weak three dimensional coupling that disrupts the scaling  \cite{mk1,kap1}. 

%\section{Conclusions}
The series of similarities between the HTSC and InOx data may point to a new understanding of the normal-state phase found in the cuprates for high fields. What was previously denoted the ``normal state" is more likely the development of a Bose-insulating phase in which incipient superconductivity persists. The magnetic field appears to drive two competing effects: increasing the field promotes  phase-fluctuations which drive the system further Bose-insulating, while also directly depairing the Cooper pairs, thereby weakening this boson-dominated phase. The former effect was first proposed by Doniach and Inui \cite{doniach} in the context of HTSC. The latter suppresses the superconducting amplitude until superconductivity is destroyed entirely. The two mechanisms together give rise to the isotherm peak which reflects an underlying existence of $H_{c2}$ above which pairs are broken. The fact that the complete magnetoresistance peak was not observed previously in the cuprates is a consequence of their very high magnetic field scale. As we noted above, a close examination of the data of \cite{ando1} shows that above 50 T the magnetoresistance starts to decrease. Accepting that the envelope behavior and the peak in the magnetoresistance in the two systems have similar sources, we expect that for $\rm La_{2-x}Sr_xCuO_4$, increasing the magnetic field to $\sim$ 200 T will yield results similar to that seen in the inset to Fig.~\ref{env}b for InOx up to $\sim$ 30 T in which the pair-amplitude is \emph{very} slowly suppressed out to high fields and the system retains a vestige of superconductivity at magnetic fields well above $H_{c2}$. Our analysis  is completely consistent with  recent Nernst effect measurements  \cite{ong1,ong2,capan} in which a strong Nernst signal was measured at fields well above the mean field $H_{c2}$ \cite{hc2} for several cuprates including $\rm La_{2-x}Sr_xCuO_4$ and was interpreted as persisting vortices \cite{ong1}.   Finally, we note that studies of more disordered (possibly more underdoped) highly anisotropic HTSC samples  may yield stronger insulating behavior similar to that found in more disordered InOx films.\\

\noindent {\bf Acknowledgments:}   We thank Yoichi Ando and Steven Kivelson for helpful discussions, and  Tim Murphy and Eric Palm for their help at the NHMFL, funded by the NSF and the State of Florida. MAS and AK supported by NSF Grant:  NSF-DMR-0119027.


\begin{references}
\bibitem{disorder}
While disorder clearly plays a major role in InOx and is certainly present in all HTSC, we believe that the main effect of disorder relevant to this work is to suppress the superfluid density.
\bibitem{hc2}
We note that $H_{c2}$ is defined only in a mean field sense. While pairs and vortices may persist to high field, this is not an indication for a much higher $H_{c2}$ since no true phase transition occurs above the SIT as the system crossovers to a  Fermi-dominated system.
\bibitem{kohler}
Here we use $\Delta \rho/\rho_n = f(H/\rho_n)$, where $\rho_n$ is the normal state resistivity and $f(x)$ is a scaling function which for most fields in the range of interest should be approximately $f(x) \propto x^2$  \cite{kapitza}.
\end{references}
\end{document}